\documentclass[preprint,12pt]{elsarticle}
\usepackage{amssymb}
\usepackage{epsfig,graphics,graphicx,bm}
\journal{Icarus}
\begin{document}
\begin{frontmatter}
\title{Giga-Year Evolution of Jupiter Trojans and the Asymmetry Problem}
\author[1,2]{Romina P. Di Sisto}
\ead{romina@fcaglp.unlp.edu.ar}
\author[1,3]{Ximena S.  Ramos}
\author[3]{Cristi\'an Beaug\'e}
\address[1]{Facultad de Ciencias Astron\'omicas y Geof\'\i sicas Universidad Nacional de La Plata}
\address[2]{Instituto de Astrof\'{\i}sica de La Plata, CCT La Plata-CONICET-UNLP  Paseo del Bosque 
S/N (1900), La Plata, Argentina}
\address[3]{Instituto de Astronom\'{\i}a Te\'orica y Experimental (IATE), Observatorio Astron\'omico, Universidad Nacional de C\'ordoba, Laprida 854, X5000BGR C\'ordoba, Argentina }

\begin{abstract}

We present a series of numerical integrations of observed and fictitious Jupiter Trojan asteroids, under the gravitational effects of the four outer planets, for time-spans comparable with the age of the Solar System. From these results we calculate the escape rate from each Lagrange point, and construct dynamical maps of ``permanence'' time in different regions of the phase space.

Fictitious asteroids in $L_4$ and $L_5$ show no significant difference, showing almost identical dynamical maps and escape rates. For real Trojans, however, we found that approximately $23 \%$ o   f the members of the leading swarm escaped after $4.5$ Gyrs, while this number increased to $28.3 \%$ for $L_5$. This implies that the asymmetry between the two populations increases with time, indicating that it may have been smaller at the time of formation/capture of these asteroids. Nevertheless, the difference in chaotic diffusion cannot, in itself, account for the current observed asymmetry ($\sim 40 \%$), and must be primarily primordial and characteristic of the capture mechanism of the Trojans. 

Finally, we calculate new proper elements for all the numbered Trojans using the semi-analytical approach of Beaug\'e and Roig (2001), and compare the results with the numerical estimations by Bro\v{z} and Rosehnal (2011). For asteroids that were already numbered in 2011, both methods yield very similar results, while significant differences were found for those bodies that became numbered after 2011.

\end{abstract}
\end{frontmatter}
     
\section{Introduction} 
\label{intro}

Although more than a century has passed since the discovery of the first Trojan asteroid in Jupiter's orbit (Wolf 1906), the origin and orbital evolution of the Jupiter Trojans is still a matter of debate. As of March 2012, a total of 5179 members have been cataloged, including numbered and multi-oppositional asteroids. Of these, 3394 are located around $L_4$, while only 1785 inhabit the tadpole region around $L_5$.

In recent years a number of Trojans have also been detected around other planets (e.g., Innanen 1991, Tabachnik and Evans 1999, Connors et al. 2011). Although those associated to the terrestrial planets are believed to be dynamically unstable in the long run, and therefore temporary populations, the asteroids associated to the outer planets appear more long lived. In particular, there is evidence that Neptune houses a Trojan population that rivals and may even surpass that around Jupiter (Chiang and Lithwick 2005). 

Many different mechanisms have been proposed to explain the origin of these bodies, particularly the Jupiter Trojans. Traditionally, these mechanisms have either disregarded planetary migration or assumed that any variation in the orbital architecture of Jupiter and Saturn was fairly smooth and adiabatic. According to this scenario, rouge asteroids in heliocentric orbits were
trapped into the Lagrange points either through the effects of gas drag with the primordial nebula (Kary and Lissauer 1995) or through the increase in size of the tadpole regions accompanying the mass growth of Jupiter itself (Marzari and Scholl 1998). Collisions among these asteroids could also have caused sufficient changes in their orbital elements to cause permanent trapping around the equilateral Lagrange points (Shoemaker et al. 1989).

Gomes (1998) and Michtchenko et al. (2001) analyzed the stability of the Trojan region assuming that Jupiter and Saturn were locked in mean-motion resonances (MMR). They found that the tadpole region would then become unstable, ejecting any primordial Trojan there in place. Although the aim of these papers was to place limits on planetary migration, Morbidelli et al. (2005) inverted this interpretation and pointed out that the same instability could also allow for the capture of new asteroids into the region. This new scenario, dubbed {\it chaotic capture}, appeared as a natural consequence of the chaotic evolution of the giant planets under the Nice model. Contrary to more traditional theories, it seemed to be able to reproduce the inclination distribution, a dynamical characteristic until then elusive.

As the Nice model evolved, so did its interpretation of the origin of Trojans. Nesvorn\'y et al. (2013) presented new numerical simulations within the Jumping-Jupiter version of the Nice model
(Morbidelli et al. 2009, Nesvorn\'y 2011, Nesvorn\'y and Morbidelli 2012). This new {\it jump-capture} mechanism proposes that part of the remnant planetesimal disk was trapped in the Lagrange points following almost instantaneous changes in the semimajor axis of Jupiter caused by close encounters with ice giants. 

Perhaps the most prominent and curious dynamical characteristic of the Jupiter Trojans is the observed asymmetry between the populations in $L_4$ and $L_5$. Not only does the leading swarm have almost $40 \%$ more asteroids than the trailing region (Grav et al. 2011, Nesvorn\'y et al. 2013), but there are also significant differences in the asteroid families. While $L_4$ hosts several numerous family candidates (Eurypides being the most notorious, see Beaug\'e and Roig 2001, Bro\v{z} and Rosehnal 2011), the region around $L_5$ only contains small (albeit compact) agglomerations.

The origin of this asymmetry is still a mystery. Dynamical studies of the Trojan region show the same resonance structure and stability limits in both Lagrange points, even when considering the perturbations of additional planets (e.g. \'Erdi 1996, Marzari et al. 2002, Robutel and Gabern 2006). Most of the proposed formation mechanisms also predict similar populations in both equilateral Lagrange points, including the first versions of the Nice model (e.g. Morbidelli et al. 2005). So, it appears that even under the most complex scenarios, both $L_4$ and $L_5$ are dynamically equivalent. However, recently Hou et al. (2014) showed that a temporary asymmetry my be obtained with the same initial conditions in both tadpole regions. this asymmetry, however, is short-lived and cannot at present account for the observed disparity.

Perhaps even more drastic measures are necessary to create an asymmetry. In the mechanism proposed by Nesvorn\'y et al. (2013), close encounters with an ice giant could have partially depleted one of the Lagrange points while leaving the other virtually unaffected. Once again, as it occurs several times in exoplanetary systems, planetary scattering appears as an excitation mechanism much more effective than slow-acting long-range gravitational perturbations. Since scattering is stochastic and extremely sensitive to initial conditions, the final ratio of Trojans in $L_4$ and $L_5$ (i.e. $N(L_4)/N(L_5)$) is not deterministic. However, some of the runs presented in Nesvorn\'y et al. (2013) do seem to be able to obtain values similar to those observed in the real asteroids.

The aims of this paper are very simple. Since it is known that even today the Trojan population is undergoing slow chaotic diffusion (Tsiganis et al. 2005, \'Erdi et al. 2013), what dynamical characteristics can be considered primordial? In particular, can the current $N(L_4)/N(L_5)$ ratio be considered invariant in time, or was the original asymmetry different?  

This paper is organized as follows. In Section 2 we review and analyze the main physical and dynamical characteristics of the Trojan swarm and present the results of a long term integration of observed Trojans. In Section 3, we extend our Gyr-simulations to fictitious massless particles in $L_4$ and $L_5$, and compare those results with the evolution of the real asteroids. Finally, discussions and conclusions close the paper in Section 4.

\section{The Observed Population}

\subsection{Orbital and Dynamical Features}
\label{ini}

As of March 2013, there were 2972 numbered Jupiter Trojans, thus with fairly reliable orbits. Of these, 1975 (over $66 \%$) display tadpole orbits around $L_4$, while 997 are associated to $L_5$. The population of Jupiter Trojans is believed to be complete up to absolute magnitude $H = 12$ (Szab\'o et al. 2007); however for the purposes of the present study we will consider the complete (numbered) population regardless of the absolute magnitude. 

\begin{figure}[t!]
\centerline{\includegraphics*[width=0.9\textwidth]{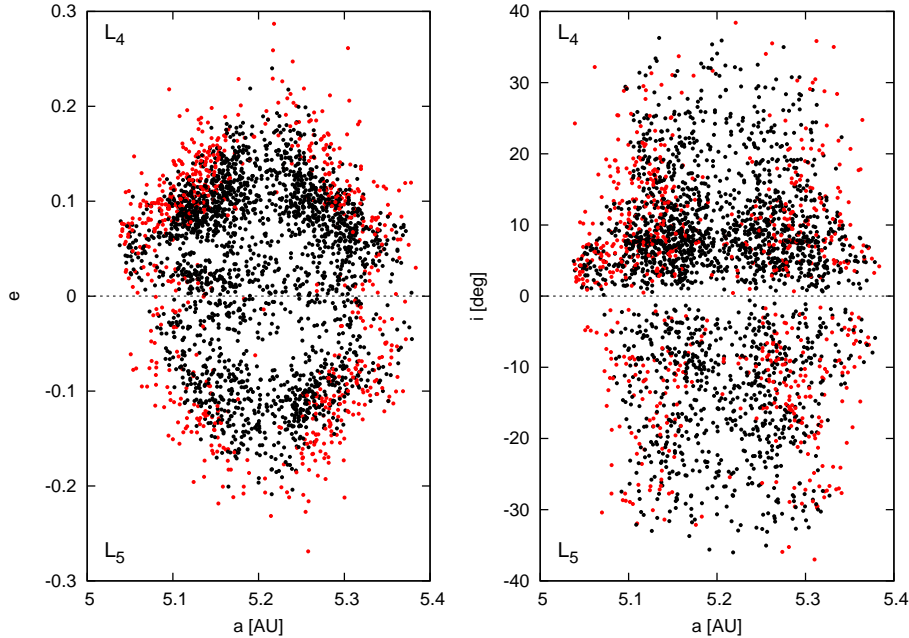}}
\caption{Distribution of eccentricity (left) and inclination (right) with the semimajor axis of all numbered Trojans in $L_4$ (top) and $L_5$ (bottom). Orbits stable for 4.5 Gyrs are shown in black, while unstable asteroids are depicted in red.}
\label{fig1}
\end{figure}

Fig. \ref{fig1} shows the distribution of both swarms in the $(e,a)$ and $(i,a)$ planes, where $a$ is the semimajor axis (in AU), $e$ the eccentricity and $i$ the inclination with respect to the Laplace plane of the outer Solar System. The upper half of the plots (positive values of $e$ and $i$) corresponds to $L_4$, while the lower half corresponds to $L_5$. The orbital elements are osculating, but each asteroid was integrated until it crossed the representative plane defined by the conditions $M - M_{\rm J} = 0$, $\varpi - \varpi_{\rm J} = \pm 60^\circ$ and $\Omega - \Omega_{\rm J}=0$. Here $M$ is the mean anomaly, $\varpi$ the longitude of pericenter and $\Omega$ the longitude of the ascending node. Variables with subscript ``J'' correspond to Jupiter. 
The orbits were evolved using the hybrid integrator EVORB (Fern\'andez et al. 2002) including the gravitational perturbations of all outer planets. The masses of the inner planets were added to the Sun, and we adopted a time-step of 0.2 years. 

One of the most interesting dynamical characteristics of the Trojan asteroids is that not all of them lie in orbits that are stable over time-spans comparable with the age of the Solar System. Although the chaotic nature of some of these asteroids has been known for many years (e.g. Milani 1993), at first it was not clear whether this chaoticity was local (i.e. ``stable-chaos'') or whether it could lead to ultimate escapes from the Lagrange points. Levison et al. (1997) were the first to present Gyr-long numerical simulations of known and fictitious Trojans, showing that indeed approximately $12 \%$ of the asteroids were unstable due to the gravitational perturbations of the other giant planets in times of the order of the age of the Solar System. Furthermore, they showed that the orbits of the escaped asteroids resemble those of the Jupiter Family Comets. 

Tsiganis et al. (2005) revisited this problem, calculating dynamical maps of Lyapunov characteristic exponents for grids of elements $(D,e)$ for a set of discrete values of the inclination $i$. Here $D$ is the semi-amplitude of libration of the asteroid. Although their total integration time was only equal to $4$ Myr, it was sufficient to correlate their maps with the distribution of real Trojans, and identify which asteroids could lie in unstable orbits. Those candidates were integrated a second time for $4.5$ Gyr, confirming the unstable nature of their motion. The results of Tsiganis et al. (2005) show that $\sim 17 \%$ of the real Trojans escape from the Lagrange regions in this time interval and are effectively unstable. The ``effective'' stability region shrinks with increasing orbital inclination.

\begin{figure}[t!]
\centerline{\includegraphics*[width=0.9\textwidth]{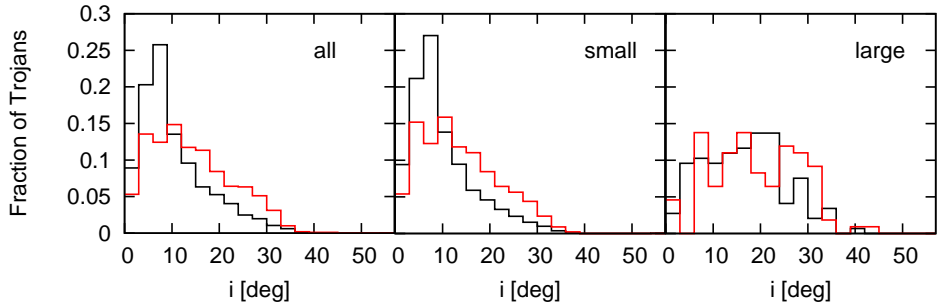}}
\caption{Distribution of inclination of all (left), small with $d<30$ (middle) and large with $d>30$ km (right) numbered $L_4$ (black) and $L_5$ (red) Trojans.}
\label{fig2}
\end{figure}

It is also possible to observe some special features of the orbital elements of the Trojan population. In Fig. \ref{fig1}, the inclinations of $L_5$ Trojans  seems to be more disperse than in $L_4$. There is a well-defined set of low inclination Trojans in $L_4$ that is not observed in such a number in $L_5$. In fact, while the mean values of osculating semimajor axis and eccentricity are almost the same for $L_4$ and $L_5$, i.e. $\langle a_{L_4} \rangle=5.2062$ AU, $\langle a_{L_5} \rangle=5.2068$ AU, $\langle e_{L_4} \rangle=0.072 , \langle e_{L_5} \rangle=0.074 $, the  mean value of the inclinations in $L_5$ is greater than in $L_4$: $\langle i_{L_5} \rangle=14^{\circ}.2, \langle i_{L_4} \rangle=10^{\circ}.4 $. Both results are in agreement with Slyusarev (2013).

However, the difference in inclination distribution is size-dependent. There are a number of papers that analyzed the dependence of the cumulative size distribution (CSD) and albedos on the Trojan sizes. Jewitt et al. (2000) found that there must be a break in the CSD at diameters $d \sim 60 - 80$ km. 
 Yoshida and Nakamura (2008) analized the CSD of $L_4$ and $L_5$ and found that on a range 
of $5$ km $< d < 93$ km, the slope of the CSD is nearly constant, breaking at $d \sim 90$ km. 
 Fraser et al. (2014) obtained a CSD power-law for Trojans that breaks at absolute magnitude
$H = 8.4$, that corresponds to a diameter $d =130$ km (for albedo 0.045). 
Fernandez et al. (2003) derived visual albedos for 32 Trojans with diameters $d > 50$ km and found a mean value of 0.056 and 0.041 depending on the beaming parameter. Later, Fernandez et al. (2009) presented thermal observations of 44 small 
Trojans with diameters $ 5 < d < 24 $ km and found a median value of 0.12, higher 
than that of the large Trojans. They attributed this correlation albedo-size 
to the collisional evolution, which makes that the smaller Trojans are more 
likely to be collisional fragments of larger bodies and thus they have younger 
surfaces implying cleanest ones. Then, considering the albedos
it seems that there is a certain diameter transition in the range  $30 < d < 50$ that divide two ``species'' of Trojans, the greater ones would
be primordial and the smaller the product of a collisional evolution and fragments of families. Considering 
the break of the CSD obtained in the previous mentioned papers, a transition between small and large Trojans would be in the range $60 < d < 130$. 
    Taking those works into account, we chose a transition diameter between ``small'' and ``large'' Trojans at $d_t = 30$ km  in order to analyze the inclination distribution size dependence. We test however,  
 greater values of $d_t$   which showed  no significant differences.
We have calculated that  $\langle i_{L_5} (d<30$ km$) \rangle =13^{\circ}.2$ meanwhile  $\langle i_{L_4} (d<30$ km$) \rangle =9^{\circ}.6 $ and  $\langle i_{L_5} (d>30$ km$) \rangle =18^{\circ}$ meanwhile  $\langle i_{L_4} (d>30$ km$) =15^{\circ} $. In Fig. \ref{fig2} we plot the normalized distribution of $L_4$ and $L_5$ Trojans for all of them and for the smaller ($d<30$ km) and the greater ones ($d>30$ km). We can see that the $L_4$ population has a set of low inclination Trojans proportionally more numerous than the $L_5$ population. This trend is maintained for the smaller Trojans but it is not noticed for the greater ones. This means that there are more $L_4$ small Trojans in the  small inclination population (say less than $10^{\circ}$) than $L_5$ small Trojans. However the whole $L_5$ population is ``biased'' to high inclinations with respect to the $L_4$ population and this behavior is present for both the smaller and the greater ones. The question is to discern if this is primordial or a consequence of a different evolution? We will address this topic again in the future sections.

Contrary to the main belt asteroids, where chaotic diffusion is mainly fueled by non-conservative forces like  Yarkovsky thermal effects (e.g. Farinella and Vokrouhlick\'y 1999), the inherent instability of the Trojans appears purely gravitational and caused by secondary and secular resonances within the tadpole regions (Robutel et al. 2005, Robutel and Gabern 2006). This seems to indicate that the full orbital characteristics of the Trojans cannot be estimated from short (Myr) timescale integrations, and that very slow diffusive and dynamical effects contribute to sculpt the distribution of these swarms. 

The implications of the results of Levison et al. (1997) and Tsiganis et al. (2005) are sufficiently interesting to merit a new analysis. Contrary to these works, we analyzed each 
Lagrange point separately, analyzing whether the instability of each swarm showed differences.

\subsection{Long-term Orbital Evolution}
\label{sim}

We performed a numerical integration of all 2972 numbered Jupiter Trojans under the gravitational action of the Sun and the four giant planets over 4.5 Gyr. We used the hybrid integrator EVORB (Fern\'andez et al. 2002). Each integration was stopped if the particle suffered a close encounter with a planet (i.e. mutual distance smaller than the corresponding Hill radius). We took notice of the time of each ``escape'' and the condition under which it occurred. 

\begin{figure}[t!]
\centerline{\includegraphics*[width=0.9\textwidth]{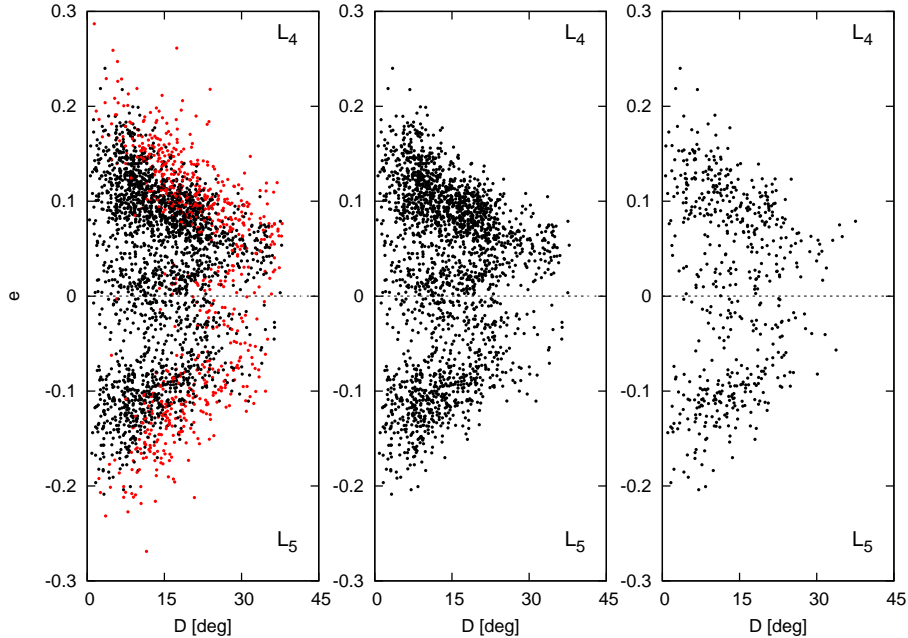}}
\caption{Distribution of eccentricity with the libration amplitude $D$ for numbered Trojans in $L_4$ (top) and $L_5$ (bottom). The left frame shows all asteroids, while the middle plot presents only those that remain stable after 4.5 Gyrs. The right-hand graph shows stable numbered Trojans with absolute magnitude $H \le 12$.}
\label{fig3}
\end{figure}

From the present day Trojans, approximately $23 \%$ were found to escape from the $L_4$ swarm, while this number increased to $28.3 \%$ for $L_5$. The percentage of $L_4$ escapees is higher than that estimated by Tsiganis et al. (2005) and about twice the proportion mentioned by Levison et al. (1997). Since our integrations included all the numbered asteroids, we believe our results are more representative of the real diffusion within the Trojan swarms. In Figures \ref{fig1} and \ref{fig3} the unstable asteroids are shown in red, while those that remained tied to tadpole orbits are depicted in black. While there appears to be little correlation between both sub-sets in the $(i,a)$ plane, the $(e,a)$ and $(e,D)$ planes show that the escaped asteroids are preferably those with larger libration amplitude or with larger orbital eccentricity. Since the Trojan population is already several Gyrs old, it is safe to assume that the original population covered a much larger region of the phase space, and that the 
escapees detected here are characterized by only very weak chaoticity. 

Approximately $99 \%$ for $L_5$ and $96 \%$ for $L_4$  of the ejections occurred after a close encounter with Jupiter and the rest due to encounters with Saturn. 

The middle graph of Figure \ref{fig3} shows the distribution of the $L_4$ and $L_5$ swarms considering only stable orbits, while the plot on the right is further restricted to those asteroids with absolute magnitude $H \le 12$. At least from a first-hand analysis, there appears no distinction between the two swarms. In all cases, however, there is a noticeable paucity of asteroids with small amplitudes of libration $D$ as well as small values of $|e-e_J|$. This is indicative of a well known property: most of the Trojans display significant amplitudes of oscillation of both the resonant and secular degrees of freedom. 

\begin{figure}[t!]
\centerline{\includegraphics*[width=0.9\textwidth]{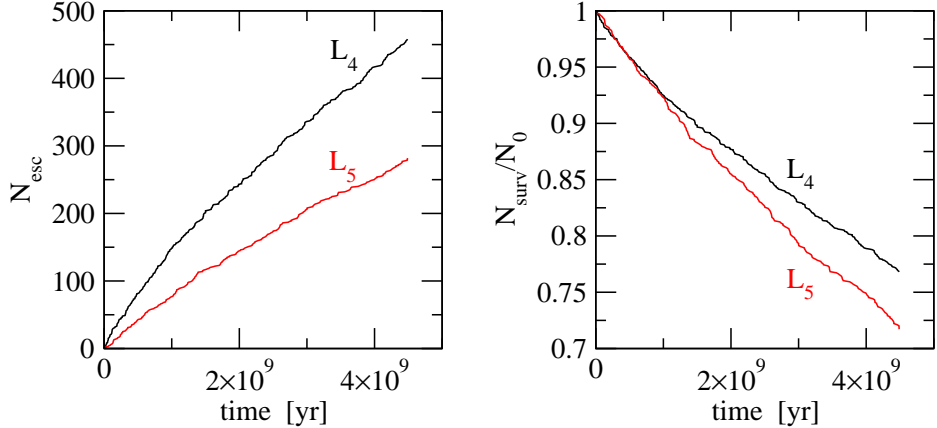}}
\caption{{\it Left:} Number of escaped numbered asteroids ($N_{\rm esc}$) as function of time, for both Trojan swarms. {\it Right:} Ratio of surviving members ($N_{\rm surv}$) over original number ($N_0$) as function of time.}
\label{fig4}
\end{figure}

A more detailed differential analysis between the two Lagrange points is shown in Figure \ref{fig4}. The left-hand graph shows the number of escapees ($N_{\rm esc}$) as a function of time, where the black curve corresponds to the $L_4$ swarm and red is reserved for $L_5$. In both cases the escape rate decreases with time until it appears close to constant for $t > 2$ Gyrs. It is very probable that the escape rate $dN_{\rm esc}/dt$ was much larger during the early stages of of the Solar System, and we are currently just detecting the tail of the distribution. However, it is interesting to note that there is no evidence of a leveling off and a tendency for $dN_{\rm esc}/dt \rightarrow 0$ at any given time. 

\begin{figure}[t!]
\centerline{\includegraphics*[width=0.6\textwidth]{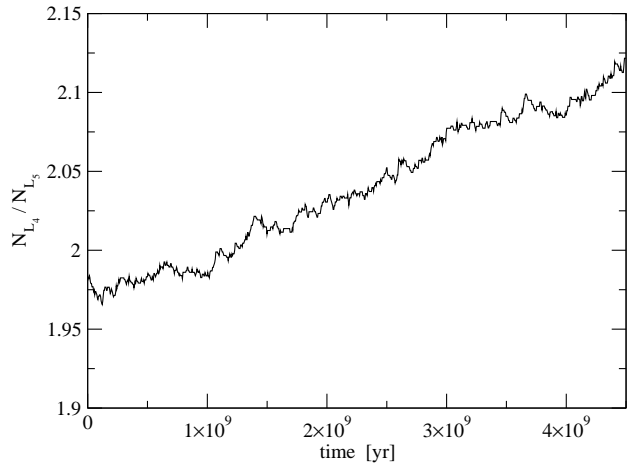}}
\caption{Time evolution of the ratio between numbered Trojans in $L_4$ and $L_5$.}
\label{fig5}
\end{figure}

While the escape rate in $L_4$ is about $\sim 60 \%$ larger than that in $L_5$, this number is affected by the larger population in the leading Trojan swarm. The right plot of Figure \ref{fig4} shows the evolution of the total number of surviving bodies in terms of the original population. As the escape rate, for $t >2$ Gyrs, the survival rate (in terms of the original population, $N_0$) $d(N_{surv}/N_0)/dt $ is almost linear. In fact it is possible to fit a linear relation for both Lagrangian points obtaining that the survival rate for $t>2$ Gyrs is given by:
$ d(N_{surv}/N_0)/dt_{L_4} = s_{L_4} = -3.9929 \times 10^{-11} \pm 2.45 \times 10^{-13} $ for $L_4$, and
$d(N_{surv}/N_0)/dt_{L_5} = s_ {L_5} = -5.11663 \times 10^{-11} \pm 4.14 \times 10^{-13} $ for $L_5$. That is, the number of surviving observed Trojans decreases with time at a rate given by $|s_{L_4}|$ and $|s_{L_5}|$ for $L_4$ and $L_5$ respectively. Then, although the trailing Trojans loose a smaller number of asteroids per unit time, the total population actually decreases faster than that associated to the leading swarm (e.g for $t>2$ Gyrs,  $ |s_{L_5}| > |s_{L_4}|$).  In consequence, the ratio between $N_{L_4}$ and $N_{L_5}$ increases with time (see Figure \ref{fig5}), from a current value of $\sim 1.97$ to a future value of $\sim 2.13$ in $4.5$ Gyrs time. Even though the two numbers are not drastically different, it is interesting to speculate whether the value could in fact have been significantly lower at the time of the capture of the Trojan asteroids. 

\begin{figure}[ht]
\centering
\begin{tabular}{cc}
\includegraphics[width=0.95\textwidth]{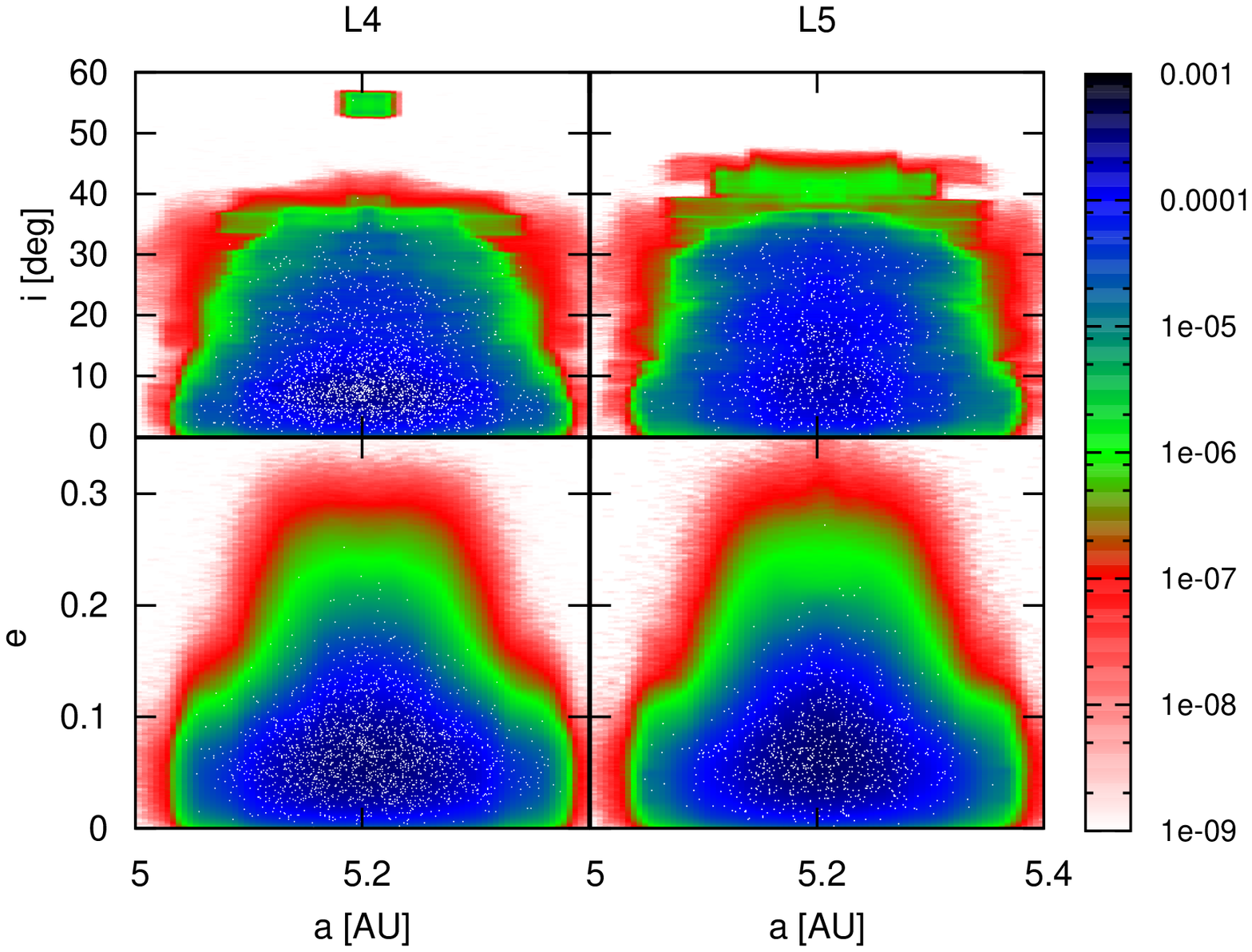} 
\end{tabular}
\caption{Normalized time-weighted distribution of the $4.5$ Gyr evolution of all numbered Trojans in the $(a,e)$ plane (left) and $(i,a)$ plane (right) for $L_4$ and $L_5$. The white dots are the numbered Jupiter Trojans.}
\label{fig6}
\end{figure}

A different way to visualize the long-term evolution is depicted in Figure \ref{fig6}. Each plot shows the normalized time fraction spent by members of the $L_4$ and $L_5$ swarms in different regions of the $(a,e)$ and $(a,i)$ planes. The color code is indicative of the portion of time  or permanence time spent in each zone (blue for most visited regions, red for least visited). The red dots are the numbered Jupiter Trojans that are our initial conditions for the numerical simulation.
Those plots form, then dynamical maps of ``permanence''. It is important to note that those plots represent the regions of stability due to  the evolution of the observed population and they may not be equal to the ``ideal'' stability regions of the restricted three body problem. 

Although there is practically no difference between the Lagrangian points in the $(a,e)$ plane, there are in the $(a,i)$ plane different regions visited by Trojans in $L_4$ and $L_5$.  In particular there is a colored region in the the zone of inclinations between $50^{\circ}$ and $60^{\circ}$, present in the $L_4$ $a$ vs $i$ plot that it is not present in the $L_5$ plot. This is due to evolution of the stable $L_4$ Trojan (83983) or 2002 GE39  with an initial high inclination of $55^{\circ}.4$. This asteroid evolves in this high inclination zone for all the simulation time i.e. 4.5 Gyrs having at the end of the integration an inclination $i = 53^{\circ}.71$.  Also the colored zones in $L_4$ reach $40^{\circ}$ and then there is a gap up to the zone of (83983). But in $L_5$ there are two asteroids (19844) and (12999) with inclinations between $40^{\circ}$ and $50^{\circ}$ that evolve in that region and escape the swarm due to an encounter with Jupiter at almost the end of the 4.5 Gyr integration.     

\begin{figure}[t!]
\centerline{\includegraphics*[width=0.9\textwidth]{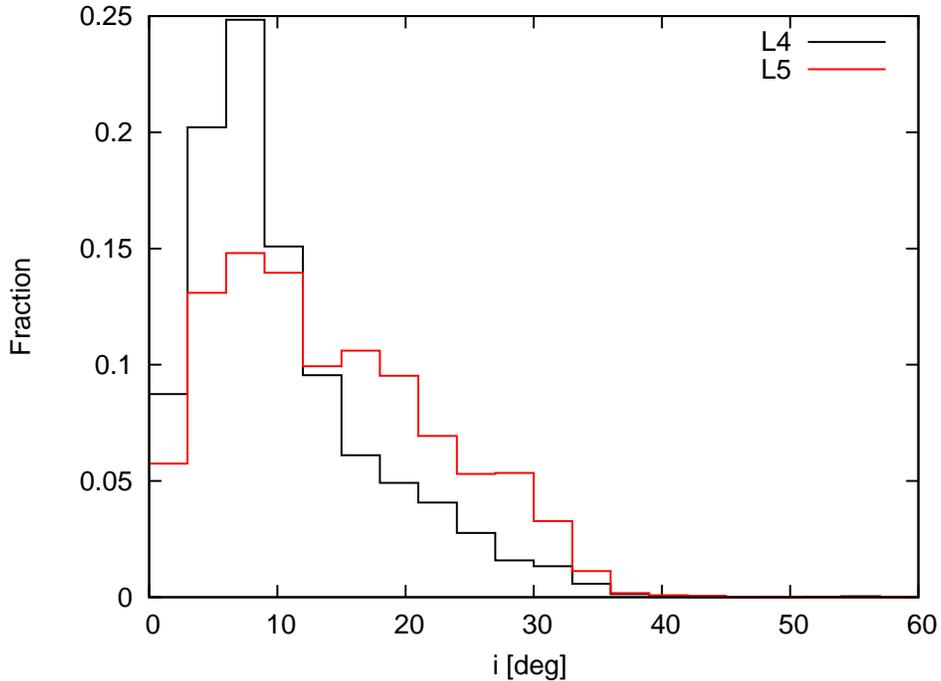}}
\caption{Normalized time weighted inclination distribution of the 4.5 Gyr evolution  
 of $L_4$ (black) and $L_5$ (red) numbered Trojans.}
\label{fig7}
\end{figure}

In line with the analysis done for the initial observed population of Trojans, we plot the normalized time weighted inclination distribution of the 4.5 Gyr evolution of the numbered Trojans,  in  Figure \ref{fig7}. We can observe that both inclination distributions hold  their original shape, with an excess of $L_4$ Trojans with $i \lesssim 10^{\circ}$ and an $L_5$ inclination distribution biased to high values with respect to the $L_4$ one. Then, the dynamical evolution of Trojans doesn't change their original inclination properties. That is, the inclination distribution of Trojans must be primordial or at least a result of an initial collisional evolution (de El\'{\i}a and Brunini (2007) and  de El\'{\i}a personal communication).   

\subsection{Proper Elements and Families}

Possibly one of the first numerical estimations of proper elements among the Trojan asteroids is due to Bien and Schubart (1987), later extended to a larger population by Milani (1993). Although the number of known asteroids at that time was small compared to today's standards (the latter  work only considered 147 bodies), Milani (1993) already identified the existence of Trojans in chaotic orbits and postulated the existence of several agglomerations in the proper-element space. Beaug\'e and Roig (2001) extended the calculation to a larger data base employing a semi-analytical model involving averaging methods and adiabatic invariant theory. They found several possible dynamical families in $L_4$, the most prominent related to Menelaus and Eurybates. No significant accumulations were found around $L_5$. 

More recently, Bro\v{z} and Rozehnal (2011) again estimated proper elements for the Trojans using a numerical technique similar to Milani (1993). They complemented their work with a detailed analysis of the size distribution of possible agglomerations and taxonomical similarities. They concluded that the only robust family appears to be the inner core of the Menelaus family proposed by Beaug\'e and Roig (2001), whose largest member is Eurybates. Again, no significant agglomeration was found in $L_5$.

All these works, however, included both numbered and multi-oppositional asteroids. Although this
was inevitable when the population of numbered bodies was small, it appears no longer necessary and may lead to significant uncertainties in the corresponding proper elements. For our present study we only consider the 2972 numbered Trojans, and calculated their proper elements using the semi-analytical model of Beaug\'e and Roig (2001). Our first aim is twofold: (i) estimate the accuracy of this perturbation model against N-body calculations, and (ii) analyze the precision of the orbital elements of the multi-oppositional bodies compared with the numbered asteroids.

\begin{figure}[t!]
\centerline{\includegraphics*[width=0.9\textwidth]{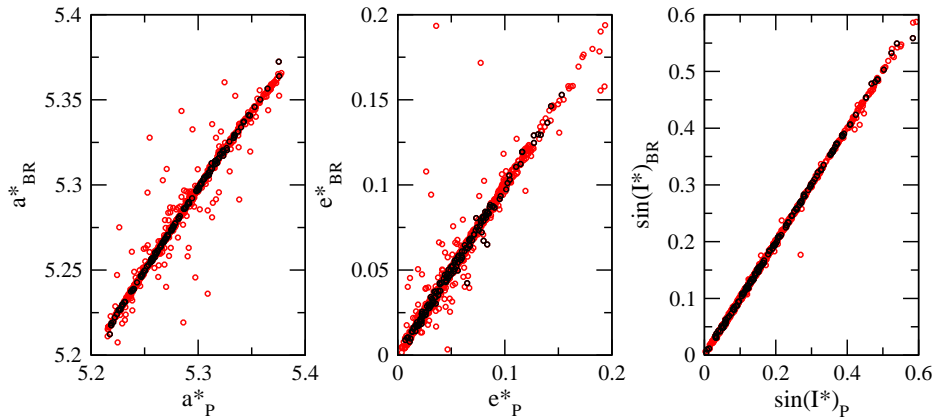}}
\caption{Comparison between the proper elements $a^*$, $e^*$ and $I^*$ from Bro\v{z} and Rozehnal (2011) (ordinates) and the values for the same asteroids estimated with the semi-analytical model of Beaug\'e and Roig (2001) (abscissas). Black circles show those bodies which where already numbered in 2011, while red circles correspond to asteroids that where updated to numbered after this time. }
\label{fig8}
\end{figure}

Results on both issues are presented in Figure \ref{fig8}, where we compare the values from the Bro\v{z} and Rozehnal catalog (ordinates) against the proper elements estimated here (abscissas). Since the definition of both sets are not exactly equal, we do not expect an exact linear trend between them, but any dispersion with respect to a one-dimensional curve is indicative of errors in the determination. 

We can see two interesting trends. First, for asteroids that were already numbered in 2011
(black circles) there is a very good agreement between both determinations. This lends credibility to the semi-analytical perturbation theory of Beaug\'e and Roig (2001), and indicates that it may be considered a good model for the long-term (albeit) regular evolution of co-orbital asteroids. Second, and more important, there is a significant dispersion in results for those asteroids which are now numbered but were multi-oppositional at the time of the work of Bro\v{z} and Rozehnal (2011). Not all of these bodies are equally unreliable, but it does show that using un-numbered asteroids for this problem may lead to highly imprecise results. 

Curiously, however, the dispersion is not significant in the proper inclination $I^*$, but is clearly noticeable in both $a^*$ and $e^*$. Thus, any proposed collisional family that includes multi-oppositional members (such as the Ennomos and 1996RJ clusters mentioned by Bro\v{z} and Rozehanl (2011)) should be considered with care.

\section{Long-term Evolution of Fictitious Trojans}

Although we detected some differences in the long-term dynamical evolution of the observed numbered Trojans, we have yet to determine whether this is due to disparity in the initial conditions (i.e. primordial) or consequence of different dynamical evolution of the $L_4$ and $L_5$ swarms. In order to address this issue, we performed a series of numerical simulations of fictitious Trojans, using the same initial distribution in both Lagrange points, and followed their evolution for 4.5 Gyrs.

\subsection{Initial Conditions}
\label{sim2}

Around both $L_4$ and $L_5$ we generated 18200 fictitious massless bodies with semimajor axes, eccentricities and inclinations in the intervals
\begin{eqnarray}
5.01 \leq \; \; a& \leq 5.4 \nonumber \\
0    \leq \; \; e& \leq 0.34 \\
0^{\circ} \leq \; \; i& \leq 60^{\circ} \nonumber
\end{eqnarray}
with separations given by $\Delta a = 0.01$, $\Delta e = 0.01$ and $\Delta i = 5^{\circ}$. The values of the semimajor axes are given in AU. The ranges considered for these elements were chosen to cover all the orbital element space attained by the observed Trojans in our previous simulations. 

The initial angular orbital elements were taken as follows. The longitude of perihelion $\varpi$ was set equal to $\varpi = \varpi_J + 60^{\circ}$ for $L_4$ and $\varpi = \varpi_J - 60^{\circ}$ for $L_5$. The longitudes of the nodes were taken equal to that of Jupiter: $\Omega = \Omega_J$, while the mean anomalies chosen as $M = M_J$. Variables with subscript $J$ correspond to Jupiter's orbit. All initial conditions were numerically integrated over $4.5$ Gyr under the gravitational effects of all four outer planets, employing the same numerical code described in the previous section.

\subsection{Results}
\label{res}

\begin{figure}[t!]
\centerline{\includegraphics*[width=0.9\textwidth]{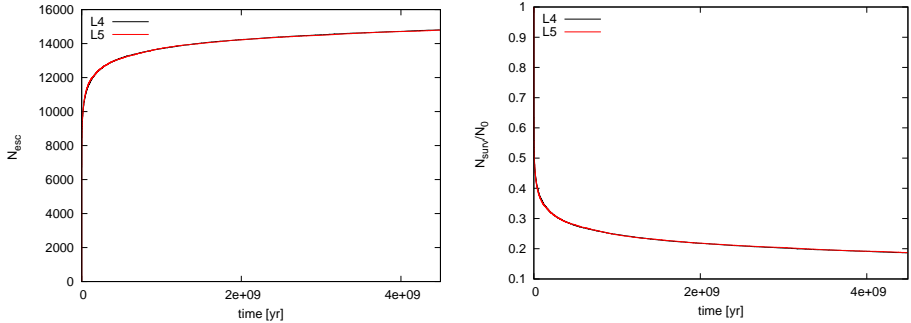}}
\caption{{\it Left:} Number of escaped fictitious Trojans ($N_{esc}$) as function of time. {\it Right:} Ratio of surviving members ($N_{\rm surv}$) over original number ($N_0 = 18200$) as function of time.}
\label{fig9}
\end{figure}

Most of the particles were ejected from the tadpole region early in the simulation, and suffered close encounters with Jupiter ($81 \%$ of the cases) or Saturn (the remaining $19 \%$). The left-hand frame of Figure \ref{fig9} shows the cumulative number of escaped objects ($N_{esc}$) as a function of time, while the graph on the right presents the ratio of surviving members ($N_{surv}$) over original number ($N_0 = 18200$). At the end of the simulation, $81.4 \%$ of the initial conditions in $L_4$ were ejected, while the number corresponding to $L_5$ was almost identical.

For both the leading and trailing populations the escape rate is almost identical throughout the integration. Almost half of the initial conditions are lost after only a few thousand years. Although the rate continues high, it slowly decreases as function of time, leading to an almost  constant value for $t > 2$ Gyrs. Fitting a straight line to the results for $t > 2$ Gyrs, we obtain that the survival rate may be approximated by:
$ d(N_{surv}/N_0)/dt_{L_4} = sf_{L_4} = -1.10675 \times 10^{-11} \pm 8.819 \times 10^{-14} $ for $L_4$, and
$d(N_{surv}/N_0)/dt_{L_5} = sf_{L_5} = -1.07656 \times 10^{-11} \pm 9.248 \times 10^{-14} $ for $L_5$. The values of both Lagrange points are virtually identical.

\begin{figure}[t!]
\centering
\begin{tabular}{cc}
\includegraphics[width=0.95\textwidth]{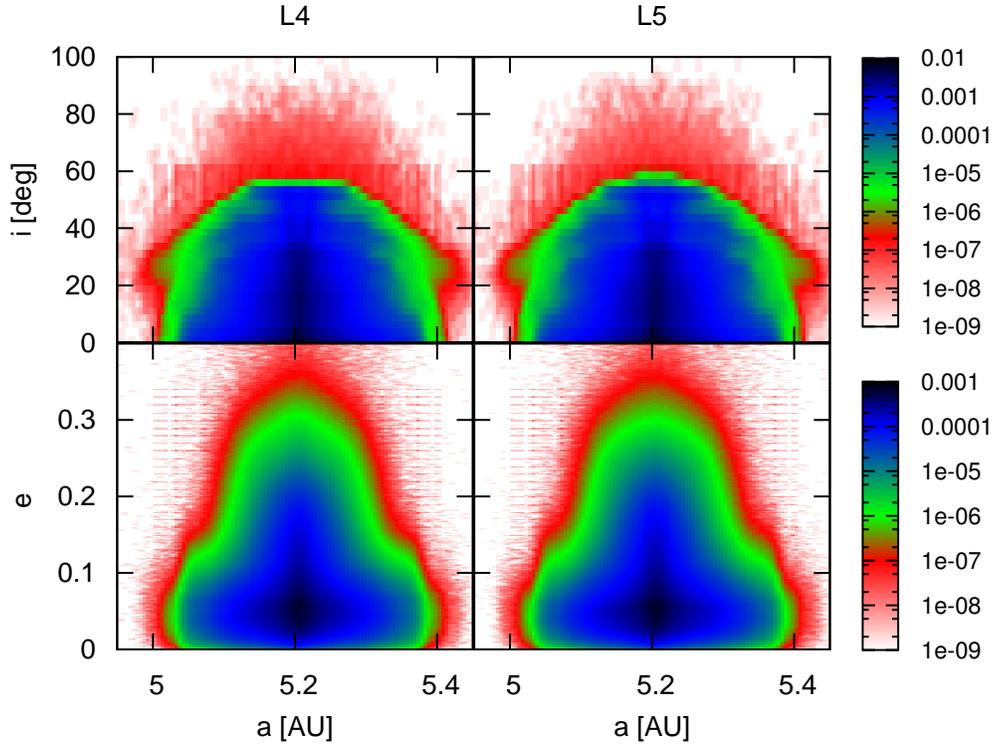} 
\end{tabular}
\caption{Normalized time-weighted distribution of the $4.5$ Gyr dynamical evolution of fictitious Trojans in the $(a,e)$ plane (left) and $(i,a)$ plane (right) for $L_4$ and $L_5$.}
\label{fig10}
\end{figure}

Figure \ref{fig10} shows the time-weighted distribution of the osculating elements of the fictitious Trojans throughout the simulation. These dynamical maps of permanence are analogous to those shown in Figure \ref{fig6}, but unaffected by the initial conditions of the particles. We find no significant difference between the two tadpole regions, indicating that not only the escape rate, but also their dynamical structure appear to be equivalent, even on Gyr timescales.

\subsection{Comparison with the Observed Population}

Even though the synthetic populations show the same escape rate, our simulations with real Trojans
indicate that the trailing population actually decreases faster than that associated to the leading swarm. At the end of their dynamical evolution, 1517 $L_4$ Trojans (out of 1975) and 715 $L_5$ asteroids (out of 997) survived in the integration. If this proportion is representative of the different escape rate of the original swarms, then we can relate the present-day populations $N_{surv} (L_i)$ with the original numbers $N_0(L_i)$ via a simple linear law:

\begin{equation}
\frac{N_{surv} (L_4)}{N_0(L_4)} \simeq 1.07 \,\, \frac{N_{surv} (L_5)}{N_0(L_5)} .
\label{ns}
\end{equation}
Even though this difference is not large, it is statistically significant when compared with the escape rates of the s
synthetic population. 

If, as calculated by Grav et al. (2011), the present unbiased fraction of Trojans between $L_4$ and $L_5$ is $N_{surv}(L_4)/N_{surv}(L_5) = 1.4 \pm 0.2$, then, from Eq. \ref{ns} we can estimate the original population ratio as

\begin{equation}
N_0(L_4)/ N_0(L_5) = 1.3 \pm 0.2 ,
\label{ni}
\end{equation}
With this calculation, the difference in the survival rate between $L_5$ and $L_4$, accounts for 
$\sim 10 \%$ of the total asymmetry or, in other words, it has contributed to $\sim 25 \%$ of the present unbiased asymmetry. Consequently, at least part of the observed asymmetry must be primordial and related to the capture/formation process of these asteroids. 

Since the dynamics appear equal, then the difference must lie in the distribution of the Trojans
around each Lagrange point. As we mentioned in Sect (\ref{ini}), the inclinations of the observed Trojans  in $L_4$ show an excess with respect to the inclination distribution of $L_5$ Trojans in the range $i \lesssim 10^{\circ}$ and this is attributed mainly to small-size bodies. However, even large-size Trojans in $L_5$ appear to have a broader inclination distribution, so the difference cannot be solely due to small asteroids. 

One possible scenario has been recently proposed by Nesvorn\'y et al. (2013), where a hypothetical ice giant planet (later ejected from the Solar System) transversed the $L_5$ cloud, partially scattering its population and modifying the inclination distribution. Another possibility could be related to different collisional evolution in each Lagrange point, for example, due to larger asteroids captured in $L_4$ than in $L_5$, producing a larger collisional cascade of fragments at  $L_4$. This could also explain the numerous family candidates at $L_4$, while only small agglomerations are detected around $L_5$.

\section{Discussion and Conclusions}

In this paper we have presented a series of numerical integrations of real and fictitious Jupiter Trojans over time-spans comparable with the age of the Solar System. We obtained that fictitious bodies with the same initial distribution in both Lagrange points, show the same dynamical evolution and orbital instability for $L_4$ and $L_5$. However the evolution of the observed population in the trailing Trojan point decays faster than that associated to the leading region. For the present day Trojans, approximately $23 \%$ were found to escape from $L_4$ swarm after $4.5$ Gyrs, while this number increased to $28.3 \%$ for $L_5$. We believe this is mainly due to a difference in the orbital element distribution of the bodies and not to any inherent dynamical process which may be more effective in one of the tadpole regions. 

From the present-day ratio of asteroids in $L_4$ with respect to $L_5$, we have estimated the original ratio, assuming that the current escape rates may be extrapolated backwards to primordial times. This assumption is not obvious, and must be considered with care. However, the results show that the original population ratio must have been much closer to unity, indicating that perhaps both Lagrange points contained originally similar number of asteroids. 

Finally, we have also calculated the proper elements of numbered asteroids using the semi-analytical method of Beaug\'e and Roig (2001), and compared the results with the numerical estimations by Bro\v{z} and Rosehnal (2011). We found that both methods yield very similar results for numbered asteroids, but not for asteroids that are now numbered but were multi-oppositional in 2011. This seems to indicate that only the numbered asteroids have sufficiently well determined orbits to allow for detailed and long-term dynamical analysis.

\vspace*{0.5cm}
\noindent{\bf Acknowledgments:}
The authors wish to express their gratitude to the reviewers for their suggestions and comments, and to the FCAGLP for extensive use of their computing facilities. This work was partially supported by research grants from CONICET and Secyt/UNC.

\end{document}